\documentclass[aps,prd,twocolumn,amsmath,showpacs,superscriptaddress,nofootinbib]{revtex4-1}

\usepackage{color}
\usepackage{graphicx}
\usepackage{dcolumn}
\usepackage{bm}

\begin{document}

\preprint{GCON-2018/05}

\title{Physical constraints on interacting dark energy models}

\author{J. E. Gonzalez}
\email{javierernesto@on.br}
\affiliation{Departamento de F\'{\i}sica, Universidade Federal de Sergipe, 49100-000, Aracaju - SE, Brasil}
\affiliation{Observat\'orio Nacional, 20921-400, Rio de Janeiro, RJ, Brasil}

\author{H. H. B. Silva}
\email{heydson.brito@ufcg.edu.br}
\affiliation{Universidade Federal de Campina Grande, 58900-000, Cajazeiras, PB, Brasil}


\author{R. Silva}
\email{raimundosilva@fisica.ufrn.br}
\affiliation{Universidade Federal do Rio Grande do Norte, 59072-970, Departamento de F\'{\i}sica, Natal, RN, Brasil}

\author{J. S. Alcaniz}
\email{alcaniz@on.br}
\affiliation{Observat\'orio Nacional, 20921-400, Rio de Janeiro, RJ, Brasil}
\affiliation{Universidade Federal do Rio Grande do Norte, 59072-970, Departamento de F\'{\i}sica, Natal, RN, Brasil}

\date{\today}

\begin{abstract}

Physical limits on the equation-of-state (EoS) parameter of a dark energy component non-minimally coupled with the dark matter field are examined in light of the second law of thermodynamics and the positiveness of entropy. Such constraints are combined with observational data sets of type Ia supernovae, baryon acoustic oscillations and the angular acoustic scale of the cosmic microwave background to impose restrictions on the behaviour of the dark matter/dark energy interaction. Considering two EoS parameterisations of the type $w = w_0 + w_a\zeta(z)$, we derive a general expression for the evolution of the dark energy density and show that the combination of thermodynamic limits and observational data provide tight bounds on the $w_0 - w_a$ parameter space.

\end{abstract}


\maketitle

\section{\label{sec:level1}Introduction}

The physical mechanism behind the late-time cosmic acceleration is currently one of the major open problems in the field of cosmology. This phenomenon has been evidenced from analysis and interpretation of different observational data sets \cite{Riess:1998cb, Perlmutter:1998np, Eisenstein:2005su, Hinshaw:2012aka, Ade:2015xua, Ata:2017dya, Alcaniz:1999kr, Lima:2003dd,Moresco:2016nqq}  and, in the context of the general relativity theory, can be explained either if one admits the existence of an exotic field, the so-called dark energy, or if the matter content of the universe is subject to dissipative processes \cite{Lima:1999rt, Chimento:2003sb} (see \cite{Sahni:1999gb, Padmanabhan:2002ji, Weinberg:2012es} for a review).

The lack of knowledge on the nature of the dark sector has motivated several approaches to unveil the physical properties of both dark matter and dark energy. In principle, a thermodynamic analysis should be relevant to constrain the behavior of these dark components or even to restrict the range of acceptable values of their  
parameters. Many approaches of this kind have been formulated in the literature (see, e.g.,~\cite{Lima04,Silva:2007fi,brevik04,bousso05,pavon06,Lima08,Pavon:2007gt,thermo} and references therein). For instance, the thermodynamics of a dark energy component described by a varying equation-of-state parameter (EoS) $\omega=\omega(a)$ with null chemical potential ($\mu=0$) was discussed in \cite{Silva12} whereas a general treatment for dark energy thermodynamics  considering a non-zero chemical potential ($\mu \neq 0$) was presented in \cite{Silva13}, generalising the results of Refs. \cite{Lima04,Silva:2007fi,Lima08,Silva12}. 
On the other hand, motivated by a possible solution of the so-called coincidence problem~\cite{Weinberg:2000yb}, interacting models of dark matter and dark energy constitute an alternative description of the dark sector which have been largely investigated~(see, e.g., \cite{cq} and references therein). This class of models are based on the premise that there is currently no known symmetry in Nature preventing a non-minimal coupling in the dark sector and, therefore, such possibility as well as its cosmological consequences must be explored. In models of this kind the non-gravitational interactions between the fluids also contribute to their density evolution, thereby  violating the usual assumption of adiabaticity (for a recent observational analysis of a large class of interacting models, see~\cite{Cid:2018ugy}).

In this paper, we extend the thermodynamic analyses of \cite{Silva12,Silva13} to a more general framework which assumes a phenomenological energy exchange between the dark energy and the cold dark matter components. Using the approach of \cite{Wang:2004cp,Alcaniz:2005dg} to obtain the interaction term, we derive the evolution of dark energy density for two equation-of-state (EoS) parameterisations of the type $w = w_0 + w_a\zeta(z)$ \cite{15a,15b,BA} and impose physical constraints on its parameters from both the second law of thermodynamics and the positiveness of entropy.  We also perform a joint statistical analysis using current observational data from distance measurements to type Ia supernovae (SNe Ia) from the JLA compilation \cite{Betoule}, measurements of $\theta(z)$ obtained from the baryon acoustic oscillations (BAO) signal using the angular two-point correlation function (2PACF)~\cite{Carvalho:2015ica,Alcaniz:2016ryy,Carvalho:2017tuu,deCarvalho:2017xye} and the angular acoustic scale of the cosmic microwave background (CMB) provided by the Planck Collaboration 2015 \cite{Ade}. In our analysis, we also use the latest measurement of the local expansion rate $H_0$, as reported in~\cite{RiessGaia}. We show that the usual constraints on the $w_0 - w_a$ parametric space are significantly enhanced when the thermodynamic bounds are incorporated in the observational analysis. Throughout this paper a subscript 0 stands for present-day quantities and a dot denotes time derivative. We assume a flat background and work with units where the speed of light  $c=1$.

\section{Interacting models}

First let us consider that the energy-momentum tensor of the cosmic
fluid $T^{\mu \nu}$ consists of two perfect fluid parts, i.e.,
\begin{equation}
T^{\mu \nu}=T^{\mu \nu}_1 + T^{\mu \nu}_2\,,\label{eq1}
\end{equation}
with $T^{\mu \nu}_i=(p_i+\rho_i)u^{\mu} u^{\nu} + p_i g^{\mu \nu}$,
where $\rho_i$ is the energy density and $p_i$ is the equilibrium
pressure of the species $i=1,2$. 
By considering the Friedmann-Lemaitre-Robertson-Walker space-time
and a coupling between these components, the condition $\nabla_{\nu}
{T}^{\mu \nu} = 0$ leads to
\begin{equation}\label{coup}
\dot{\rho}_{dm} + 3 \frac{\dot{a}}{a}\rho_{dm} = -\dot{\rho}_x -
3\frac{\dot{a}}{a}(1+ \omega)\rho_x = Q\;,
\end{equation}
where $\rho_{dm}$ and $\rho_x$ are the energy densities of cold dark
matter (DM) and dark energy (DE), respectively, while $Q$ is the
coupling function. For $Q>0$ we have the DE decaying into DM whereas for
$Q<0$ the DM component decays into DE.

In the standard context the dark matter density evolves as $\rho_{dm}
\propto a^{-3}$. However, if this component interacts with dark
energy, such interaction necessarily causes a deviation from standard evolution,
which may be characterised by the $\epsilon$ parameter, i.e.~\cite{Wang:2004cp, Alcaniz:2005dg}
\begin{equation}\label{evo}
\rho_{dm} = \rho_{dm,0} a^{-3 + \epsilon}\;,
\end{equation}
which is equivalent to a coupling term of the type 
\begin{equation} \label{acop}
Q = \epsilon H \rho_{dm}\;.
\end{equation}
where $H= \dot{a}/a$ is the Hubble parameter.  In \cite{Alcaniz:2005dg}, it was shown that the $\epsilon$ parameter must be positive, which means from Eq.~(\ref{acop}) that  $Q>0$ and, consequently, that the DE decays into DM.

For generality,  we consider that the EoS of dark energy is a function of the scale factor, $w(a)$. Replacing this into Eq. (\ref{coup}) one finds
\begin{equation}\label{evogeral}
 \rho_{x}= \frac{\tilde{\rho}_{x,0} -\epsilon \rho_{dm,0} \int{\exp
 \left[3\int{\frac{1+\omega(a)}{a}da}\right]a^{-4+\epsilon}da}}{\exp \left[3\int{\frac{1+\omega(a)}{a}da}\right]}\;,
\end{equation}
where $\tilde{\rho}_{x,0}$ is an integration constant and
$\omega=\omega(a)\equiv p_x / \rho_x$ is the time-dependent EoS parameter of dark energy fluid. In order to proceed further, we will assume the following form for the
EoS parameter: $\omega(a)=\omega_0 + \omega_a \zeta(a) $, with
$\zeta (a)$ obeying two functional forms that has been widely discussed in
the literature \cite{15a,15b,BA}\footnote{For a recent comparative study between these $w(a)$ parameterisations, we refer the reader to \cite{Escamilla-Rivera:2016aca}.}:
\begin{eqnarray}
\label{ps}
 \zeta(a) = \left\{
 \begin{tabular}{l}

${(1-a)}$  \hspace{0.65cm}  (P1)
\quad\hspace{0.73cm}\\
\\
${\frac{1-a}{2a^2-2a+1}}$  \hspace{0.36cm}  (P2)
\quad\hspace{0.19cm}\\
\end{tabular}
\right. \nonumber
\end{eqnarray}
Substituting the above parameterisations into Eq. (\ref{evogeral}) we find,
respectively,

\begin{subequations}
\begin{equation} \label{d3}
\frac{\rho^{(1)}_{x}}{\tilde{\rho}_{x,0}} = 
\frac{1-A \; \epsilon \int
a^{3(\omega_0+\omega_a)+\epsilon-1}\exp\left[3\omega_a
(1-a)\right]da}{a^{3(1+\omega_0+\omega_a)}\exp\left[3\omega_a(1-a)\right]}
\;,
\end{equation}
\begin{equation} \label{d4}
\frac{\rho^{(2)}_{x}}{\tilde{\rho}_{x,0}} =
\frac{1-A \; \epsilon \int
a^{3\omega_0+\epsilon-1}\left[\frac{a^2}{(2a^2-2a+1)}\right]^{3 \omega_a /2}da}{a^{3(1+\omega_0)}\left[\frac{a^2}{(2a^2-2a+1)}\right]^{3 \omega_a /2}}
\;,
\end{equation}
\end{subequations}
where $A \equiv \rho_{dm,0}/\tilde{\rho}_{x,0}$ is a constant.

Now, considering that the baryonic and radiation components are separately
conserved, the Friedmann equation can be written as
\begin{equation}\label{ha1}
E^{j}(z) =\left[{\Omega_{r} \over a^{4}}+{\Omega_{b} \over a^{3}} + {\Omega_{dm} \over a^{3 -\epsilon}} +
\tilde{\Omega}_{x}f^{(j)}(a)\right]^{1/2}\;,
\end{equation}
where $E^{j} = H^{j}/H_0$, the density parameters follow the usual definition, and $f^{(j)}$ stands for the ${\rho^{(f)}_{x}}/{\tilde{\rho}_{x,0}}$ ratio given by Eqs. (\ref{d3}) and (\ref{d4}). Note that the so-called dynamical $\Lambda$ models (see, e.g., \cite{Alcaniz:2012mh}) are fully recovered for values of $w_0=-1$ and $w_a=0$.

\section{Thermodynamic analysis}

In general, the thermodynamic description of the interaction between
two perfect fluids requires the knowledge of three quantities: the
energy-momentum tensor $T^{\mu \nu}_i$, given by Eq. (1), and the particle flow
vector $N^\mu_i$ and the entropy flux $S^\mu_i$ defined,
respectively, as
\begin{equation}
N^{\mu}_i = n_i u^{\mu}\,,\label{eq3}
\end{equation}
\begin{equation}
S^{\mu}_i = n_i \sigma_i u^{\mu}\,,\label{eq4}
\end{equation}
where $n_i \equiv N_i / a^3$ is the particle number density and
$\sigma_i \equiv S_i /N_i $ the specific entropy (per particle) for
each species~\cite{weinberg,silva2}. By considering that the decay into DM or DE affects only
the particle mass (the particle number is unaltered), the fluids are
composed by variable-mass particles~\cite{farrar04}. Therefore, the particle
flow vector is conserved as follows
\begin{equation}\label{eq5}
\nabla_{\mu} N^{\mu}_{i} = \dot{n}_i+ \Theta n_i =0\,,
\end{equation}
where $\Theta\equiv  \nabla_{\mu}u_{i}^{\mu}=3\dot{a}/a$ is the fluid
expansion rate. The specific entropy obeys the Gibbs equation, i.e.,
\begin{equation}\label{eq6}
n_iT_id\sigma_i = d\rho_i - {{\rho_i + p_i} \over n_i}dn_i \;,
\end{equation}
Now, assuming that $\rho_i = \rho_i(n_i,T_i)$ and $p_i = p_i(n_i,T_i)$,
it can be shown that the temperature evolution law is given by \cite{Silva12,weinberg,silva2}
\begin{equation} \label{evol-ti}
{\dot {T}_i \over T_i} = \biggl({\partial p_{0,i} \over \partial
\rho_i}\biggr)_{n_i} {\dot {n}_i \over n_i} + \biggl({\partial \Pi_i
\over
\partial \rho_i}\biggr)_{n_i} {\dot {n}_i \over n_i} \;.
\end{equation}
The fact that DM is  pressureless means that there is no
temperature evolution law for this component. Therefore, only the DE
temperature evolution law is relevant for the thermodynamic
analysis that follows. The middle and right-hand sides of Eq. (\ref{coup}), on the other hand, can
be rewritten as ${\dot \rho_x} + 3(\rho_x + p_0) \frac{\dot a} {a} =
-3 \Pi \frac{\dot a} {a}$, where we have split the dark energy
pressure into two components: $p_0=\omega_0 \rho_x$ and $\Pi$ given
by
\begin{equation} \label{visc}
 \Pi\equiv w_{a}\zeta(a)\rho_x+\frac{\epsilon}{3}\rho_{dm}\;,
\end{equation}
which mimics a fluid with bulk viscosity (see Refs. \cite{Silva12,Silva13} for a discussion). Therefore, the entropy source
of the DE fluid is \cite{silva2}
\begin{equation} \label{const1}
\nabla_{\mu} S^{\mu}_{x}=-\frac{\Pi \Theta}{T_x} \;.
\end{equation}

Considering that the DE temperature is always positive and growing in the 
course of the universe expansion (see, e.g., \cite{Lima04,Silva12,Silva13}), the second law of
thermodynamics implies that
\begin{equation}
\frac{3\omega_a \zeta(a) \rho_x}{ \epsilon \rho_{dm}}\leq -1 \;.
\end{equation}
Along with Eqs. (\ref{evo}), (\ref{d3}) and (\ref{d4}), the above inequality provides our  first thermodynamic constraint on the DE quantities. For parameterisations (P1) and (P2), they are written, respectively, as
\begin{equation} \label{c3}
\omega_a \leq -A \; \frac{\epsilon}{3}
\frac{a^{3(\omega_0+\omega_a)+\epsilon}\exp \left[ 3\omega_a(1-a)
\right](1-a)^{-1}}{\left\{ 1-A \; \epsilon \int
a^{3(\omega_0+\omega_a)+\epsilon-1}\exp\left[3\omega_a
(1-a)\right]da \right\}} \;, \nonumber
\end{equation}
\begin{equation} \label{c4}
\omega_a \leq -A \; \frac{\epsilon}{3}
\frac{a^{3\omega_0+\epsilon}\left[\frac{a^2}{2a^2-2a+1}\right]^{3 \omega_a /2}\left(\frac{1-a}{2a^2-2a+1}\right)^{-1}}{\left\{ 1-A \; \epsilon \int
a^{3\omega_0+\epsilon-1}\left[\frac{a^2}{2a^2-2a+1}\right]^{3 \omega_a /2}da \right\}} \;, \nonumber
\end{equation}
which clearly are not defined at $a=1$, where $\omega=\omega_0$. On
the other hand, using the well- known Euler relation with null
chemical potential: $T_xS_x=(\rho_x + p_x)V_x$ (where $V_x\propto
a^3$ is the comoving volume) the positiveness of entropy\footnote{As
stated by the statistical microscopic concept of entropy: $S=k_B\ln
W>0$.} requires that
\begin{equation}
[1+\omega(a)]\;\rho_x \geq 0 \;.
\end{equation}
which provides our second set of thermodynamic constraints. For parameterisations (P1) and (P2), it is written as
\begin{equation} \label{cc3}
[1+\omega_0+\omega_a (1-a)]\; \rho_x^{(1)} \geq 0 \;.
\end{equation}
\begin{equation} \label{cc4}
\left[1+\omega_0+\omega_a \frac{1-a}{2a^2-2a+1}\right]\; \rho_x^{(2)} \geq 0 \;,
\end{equation}
respectively. When the dark energy density satisfies the weak energy condition, i.e., $\rho_x \geq 0$, for all values of the scale factor $a$ in the interval of study, the second set of thermodynamic constraints is exactly equal to the one obtained for non-interacting models \cite{Silva12,Silva13}:
\begin{equation}
[1+\omega(a)] \geq 0 \;.
\end{equation}
For the case in which the dark matter and dark energy components are not coupled ($\epsilon \rightarrow 0$), one also fully recovers the results of \cite{Silva12} for the both sets of thermodynamic constraints above.

\begin{table*}
  \centering
  {

  \begin{tabular*}{\linewidth}{@{\extracolsep{\fill}}l*{9}{@{\extracolsep{\fill}}c}@{\extracolsep{\fill}}}

\hline
\hline
&$\epsilon$& $w_o $ & $w_a$ & $\Omega_{dm}$ & $H_0$ \\ 
\hline
\\
(P1)  \quad & $0.010 \pm 0.008$ & $-0.78 \pm 0.14$ & $-1.44 \pm 0.85$ & $0.223 \pm 0.012$ & $71.23\pm 1.32$ \\ 
\hline
\\
(P2)  \quad & $0.011 \pm 0.009$ & $-0.78 \pm 0.12$ & $-1.04 \pm 0.53$ & $0.228 \pm 0.012$ & $71.26\pm 1.36$ \\ 
\hline

  \end{tabular*}
}
\caption{Constraints on the cosmological parameters for P1 and P2 considering the Gaussian prior $H_0=73.52 \pm 1.62$ $\rm{km.s^{-1}.Mpc^{-1}}$ .}
  \label{tableresults}
\end{table*}

\begin{table*}
  \centering
  {

  \begin{tabular*}{\linewidth}{@{\extracolsep{\fill}}l*{9}{@{\extracolsep{\fill}}c}@{\extracolsep{\fill}}}

\hline
\hline
 \quad &$\epsilon$& $w_o $ & $w_a$ & $\Omega_{dm}$ & $H_0$\\ 
\hline
\\
(P1)  \quad & $0.072 \pm 0.046$ & $-0.83 \pm 0.16$ & $-1.21 \pm 0.99$ & $0.243 \pm 0.021$ & $64.19\pm 3.60$ \\
\hline
\\
(P2)  \quad & $0.073 \pm 0.043$ & $-0.83 \pm 0.14$ & $-0.87 \pm 0.71$ & $0.245 \pm 0.021$ & $64.19\pm 3.31$ \\ 
\hline

  \end{tabular*}
}
\caption{Constraints on the cosmological parameters for P1 and P2 considering a flat prior for each parameter.}
  \label{tableresults2}
\end{table*}

\section{Observational Data}

In order to test the class of models discussed in the previous section, we perform a Bayesian statistical ana- lysis using  different cosmological observables taking into account  the above sets of thermodynamic constraints.  

The primary data set used in this analysis is the type Ia supernovae (SNe Ia) compilation named Joint Light-curve Analysis (JLA),  which comprises  740 observational data obtained by SDSS-II and SNLS collaborations~\cite{Betoule}. The distance modulus is standardised using the model
\begin{equation}
\hat{\mu} =m_B^*-(M_B- \alpha \times  X_1 + \beta \times C )
\end{equation}
where $m_B^*$ is the observed peak magnitude in the rest frame B band, $C$ is the color at the maximum brightness, $X_1$ is the time stretching of the light-curve and $\alpha$ and $\beta$ are nuisance parameters. The absolute magnitude $M_B$ is dependent on the host galaxy properties and the effects of this dependence are corrected by the step function:

\begin{equation}
 M_B = \left\lbrace
      \begin{array}{ll}
        M^1_B &\quad \text{if} \quad  M_{\text{stellar}} < 10^{10}~M_{\odot}\,,\\
        M^1_B + \Delta_M & \quad \text{if} \quad M_{\text{stellar}} \geq 10^{10}~M_{\odot}
      \end{array}
     \right.
\end{equation}
being $M_{\text{stellar}}$ the  stellar mass of the SN host galaxy and $\Delta_M $ another nuisance calibration parameter \cite{Betoule}. The distance modulus is related to the cosmological model via the luminous distance by
\begin{equation}
 \mu_{model}(z)=5\log\left(\frac{d_L(z)}{ 1\text{Mpc}} \right) + 25,
\end{equation}
where $d_L(z)$ is the luminosity distance. Note that both the cosmological and SNe calibration parameters are fitted simultaneously.

\begin{figure*}[t]
  \includegraphics[scale=0.24]{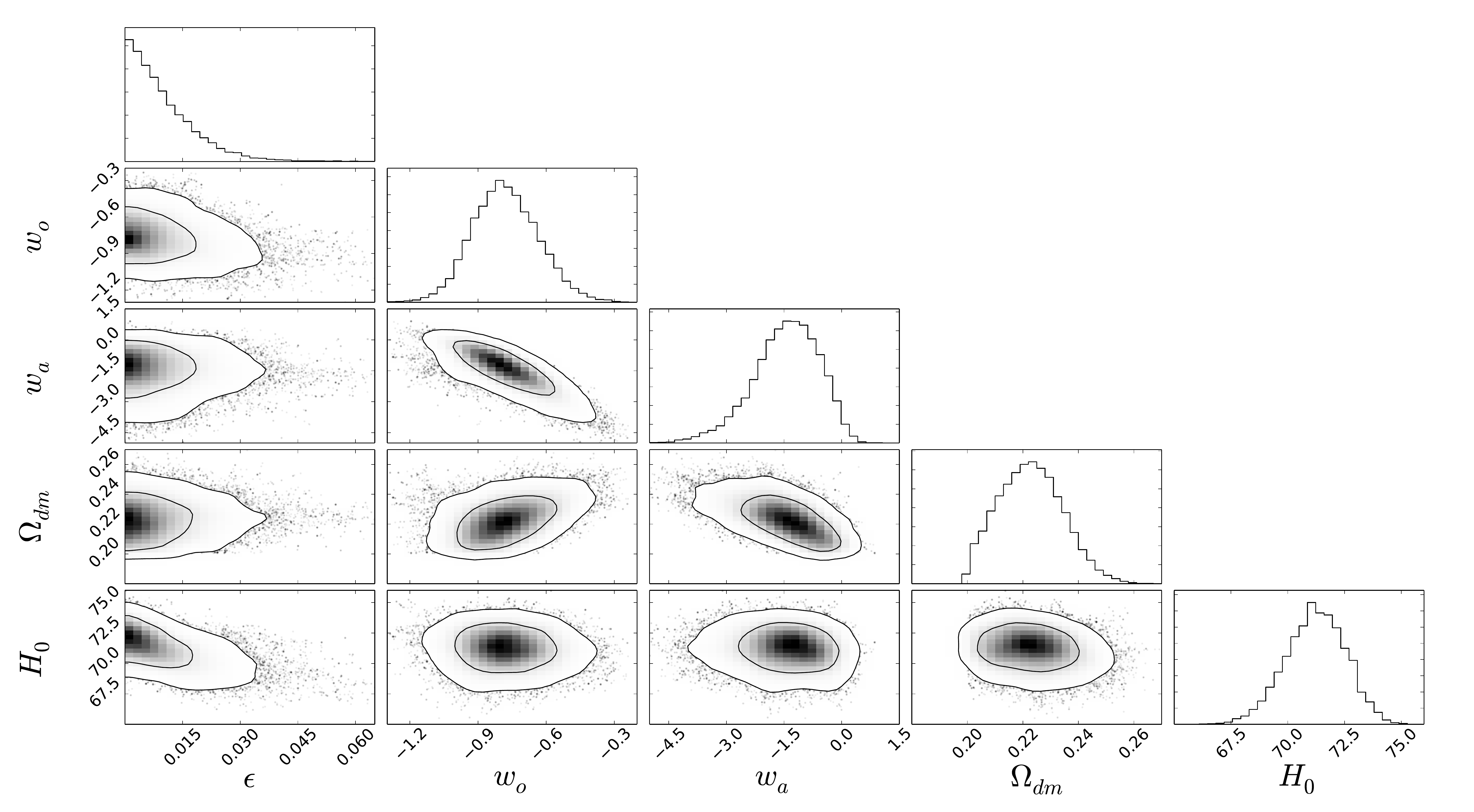}
  \caption{\small The results of our statistical analysis. Confidence contours (68.3\% and 95.4\%) and the posterior distribution for  the cosmological parameters assuming P1 and considering the Gaussian prior $H_0=73.52 \pm 1.62$ $\rm{km.s^{-1}.Mpc^{-1}}$. }
  \label{CPL5P}
\end{figure*}

\begin{figure*}[t]
  \includegraphics[scale=0.24]{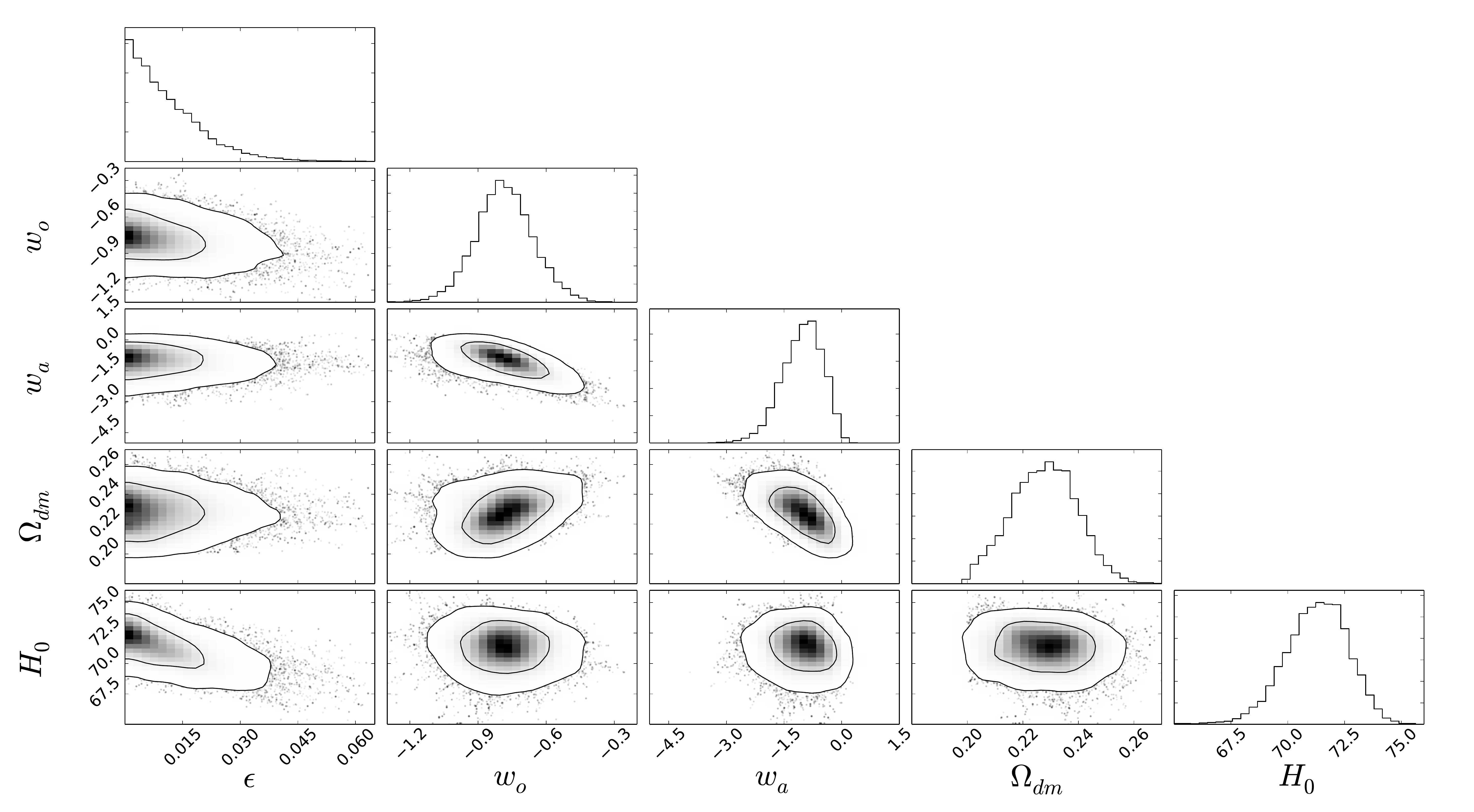}
  \caption{The same as in Figure (1) for P2.}
  \label{BA5P}
\end{figure*}

In our analysis we also use recent BAO data obtained from a 2-point angular correlation function analysis of the SDSS luminous red galaxies and quasars (hereafter $\theta_{\rm{BAO}}$)~~\cite{Carvalho:2015ica,Alcaniz:2016ryy,Carvalho:2017tuu,deCarvalho:2017xye}.  The $\theta_{\rm{BAO}}$ data are obtained by measuring the angular separation between pairs for a defined comoving acoustic scale, considering thin redshift shells of order $\delta_z =0.01 - 0.02$. Differently from the usual measurements of the BAO signal obtained from the 2-point correlation function (which assume a fiducial cosmology in
order to transform the measured angular positions and redshifts into comoving distances), the 2PACF measurements of $\theta_{\rm{BAO}}$ are almost model-independent, which makes them a robust quantity to test cosmological models. The theoretical value of $\theta_{\rm{BAO}}$ for a given cosmology is given by
\begin{equation}
\theta_{\rm{BAO}} (z) = \frac{r_s}{(1+z)d_A(z)}\;,
\end{equation}
where $d_A=d_L/(1+z)^2$ and the sound horizon scale is obtained from the expression:
\begin{equation}
 r_s(z)= \frac{1}{\sqrt{3}}\int_{z_{drag}} ^\infty \left(1+\frac{3\Omega_b}{4 \Omega_\gamma(1+z')}\right)^{-1/2}\frac{dz'}{H(z')},
\end{equation}
with $z_{drag}$ being determined by the fitting formula in \cite{Hu1996} and $\Omega_\gamma$ corresponding to the present photon density parameter. The data points used in the analysis are taken from \cite{Carvalho:2015ica,Alcaniz:2016ryy,Carvalho:2017tuu,deCarvalho:2017xye}.

Finally, we use the information of the CMB data from the Planck Collaboration encoded in the position of the first peak of the temperature power spectrum, $l_1$. The first peak at the CMB power spectrum can be calculated using the expression \cite{Hu2001}:
\begin{equation}
 l_1= l_A \left\{1-0.267\left[\frac{\rho_r(z_*)}{0.3(\rho_b(z_*)+\rho_{dm}(z_*))}\right]^{0.1}\right\},
\end{equation}
where $l_A$ is the acoustic scale given by:
\begin{equation}
 l_A=\pi (1+z_*)\frac{d_A(z_*)}{r_{\text{dec}}}.
\end{equation}
In the above expressions, $z_*$ is the decoupling redshift fitting in \cite{Hu1996} and  $r_{\text{dec}}$ is the sound horizon scale at the decoupling epoch. We use $l_1=220.0\pm 0.5$~\cite{Ade:2015xua}.

\section{Analysis and Results}

We perform a bayesian statistical analysis with the above mentioned sets of data where our posterior distribution is written in terms of the likelihood distribution, ${\cal{L}}(\Theta|d)$ and the prior distribution, $\pi(\Theta)$, as:
\begin{equation}
 P(\Theta|D) \propto {\cal{L}}(D|\Theta)\pi(\Theta),
\end{equation}
where $\Theta$ is the set of parameters and $D$ the data considered. Our  Markov Chain  Monte  Carlo (MCMC) simulations are made using the \textit{emcee} Python module \cite{Foreman2013}  assuming a Gaussian likelihood distribution,
\begin{equation}
 {\cal{L}}(D|\Theta) \propto \exp(- \chi_T^2 / 2),
\end{equation}
where the total chi-square function is the sum of the contribution of each cosmological observable, $\chi_T^2= \chi_{SNe}^2 +\chi_{BAO}^2 +\chi_{CMB}^2$. For the SNe Ia data we consider
\begin{equation}
\chi_{SNe}^2=  (\hat{ \bm{ \mu}} - {\bm \mu}_{model})^T C^{-1}_{SN}(\alpha, \beta) (\hat{\bm{\mu}} - {\bm \mu}_{model})^T.
\end{equation}
We also take into account statistical and systematic errors encoded in the SNe covariance matrix $C_{SN}(\alpha, \beta)$~\cite{Betoule}.

\begin{figure*}[t]
   \begin{center}
   \begin{tabular}{l l}
{\includegraphics*[scale=0.18]{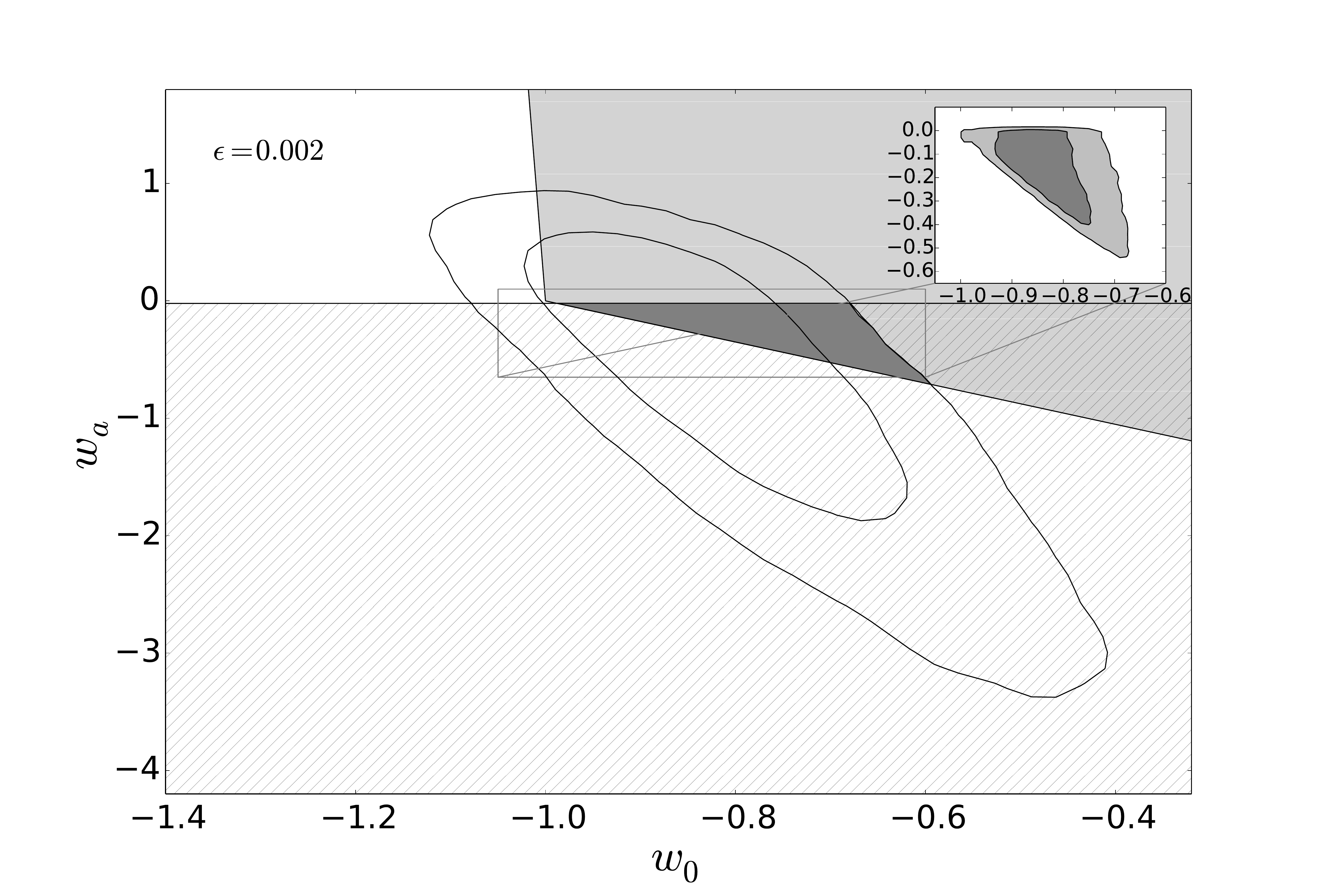}} &
{\includegraphics*[scale=0.18]{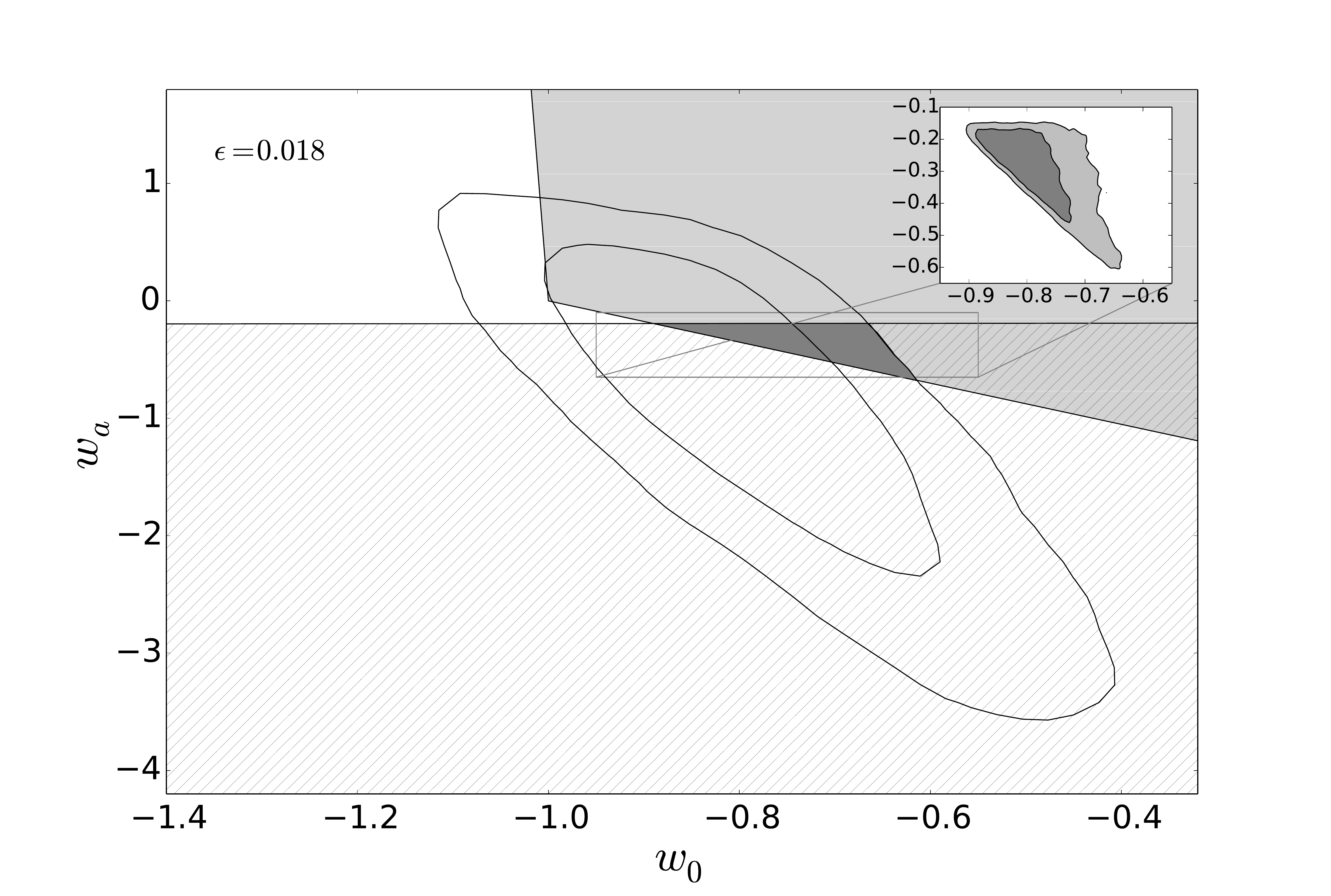}}
    \end{tabular}
  \caption{{\emph{ Left panel:}} Confidence contours (68.3\% and  the 95.4\%) on the $w_0 - w_a$ plane for parameterisation P1. The hatched and shaded regions correspond to the first and second sets of thermodynamic constraints, respectively. The confidence contours and the thermodynamic constraints assume  $\epsilon = 0.002$. {\emph{Right panel:}} The same as in the previous panel for $\epsilon = 0.018$. The dark triangle corresponds to the combined allowed region. {The subplot in each panel shows the confidence contours obtained by introducing the thermodynamical restrictions in the prior distribution $\pi(\Theta)$ to perform the statistical analysis.}}
  \label{CPLfig}
 \end{center}
\end{figure*}

\begin{figure*}[t]
   \begin{center}
   \begin{tabular}{l l}
{\includegraphics*[scale=0.18]{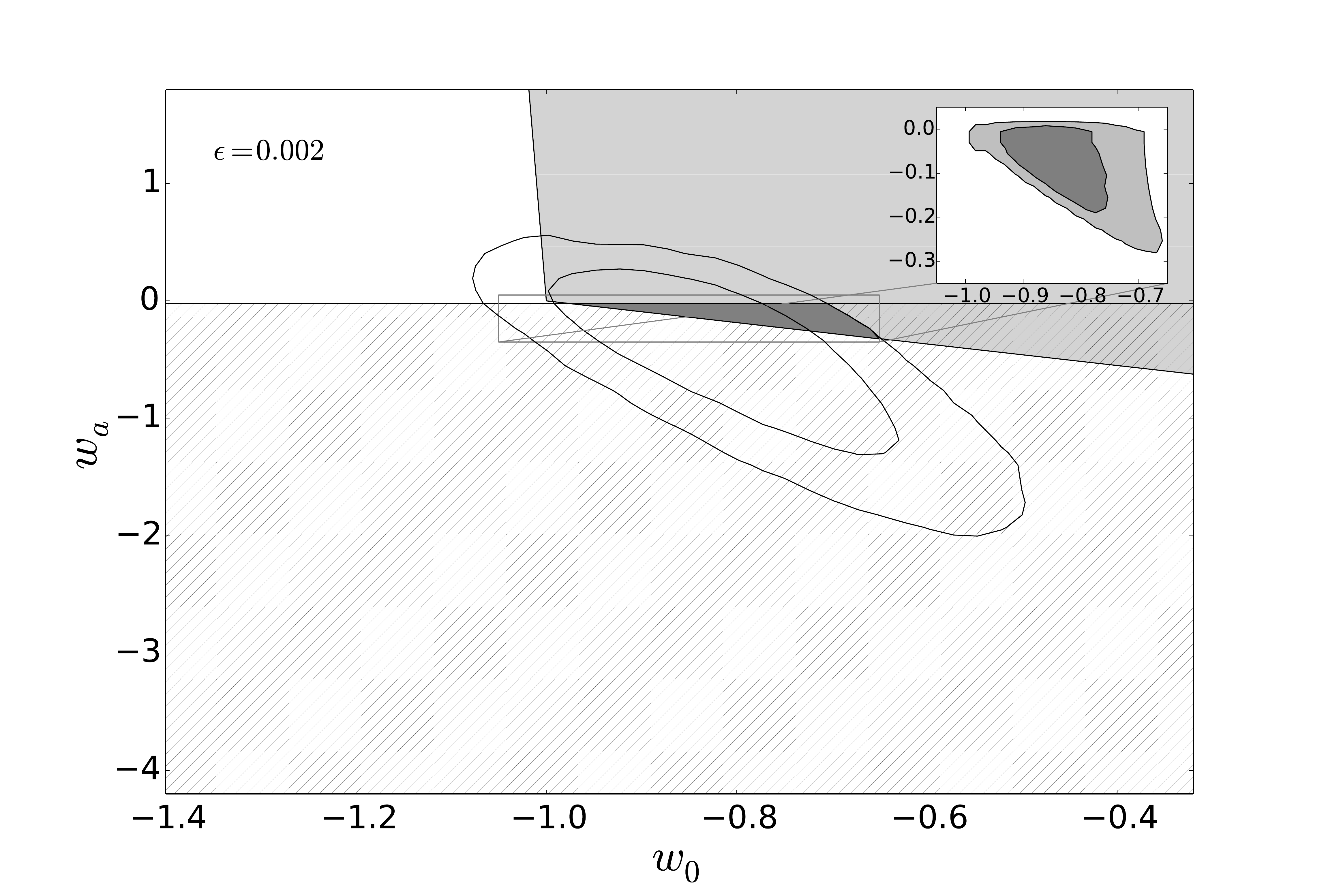}} &
{\includegraphics*[scale=0.18]{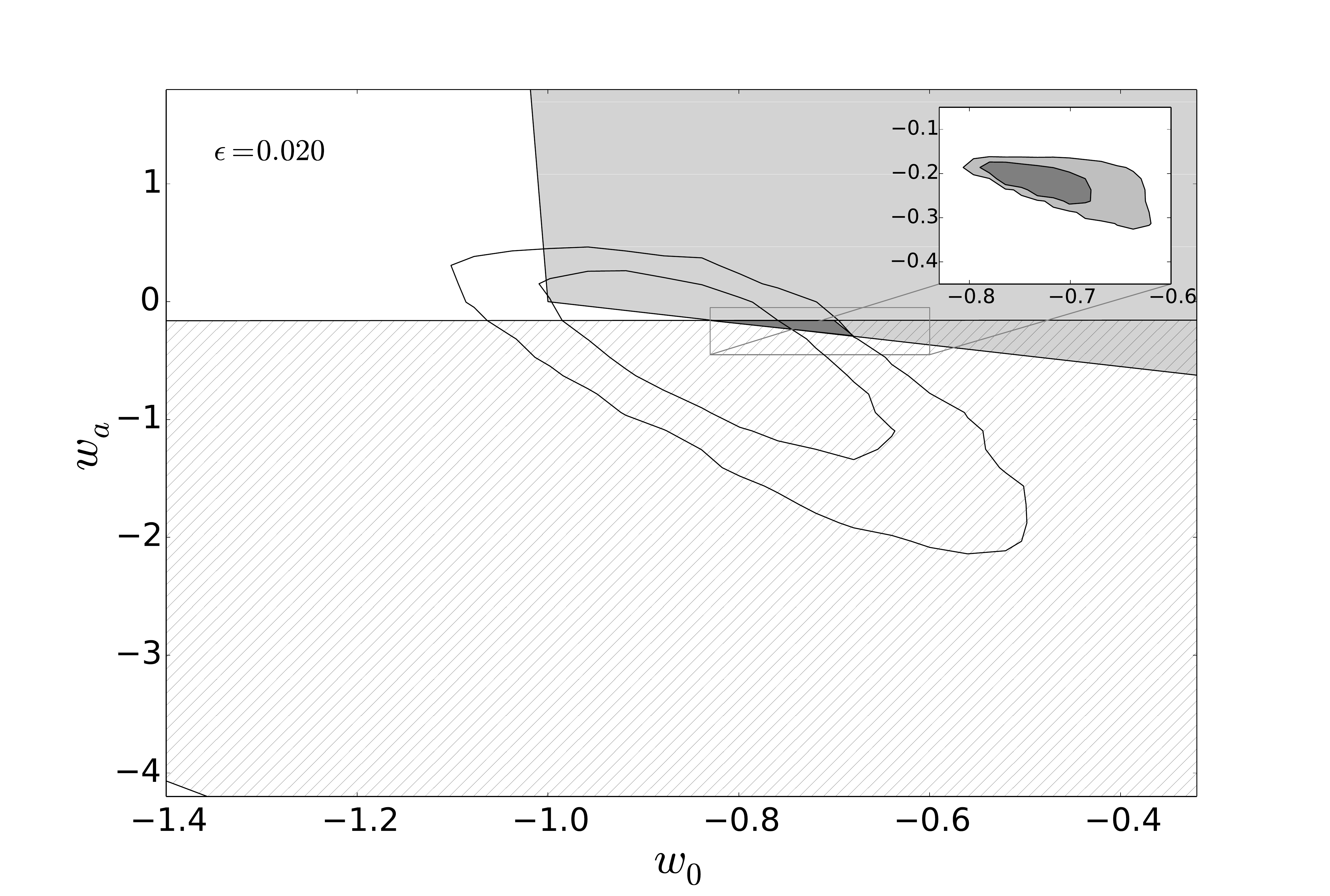}}
    \end{tabular}
  \caption{The same as in Figure (3) for P2 with $\epsilon = 0.002$ and $\epsilon = 0.02$.}
  \label{BAfig}
 \end{center}
\end{figure*}

In our statistical analysis, we use the most recent estimate of the Hubble constant $H_0=73.52\pm1.62$ $\rm{km.s^{-1}.Mpc^{-1}}$ \cite{RiessGaia} as a Gaussian prior and  flat  priors for  the other parameters. In particular, taking into account the constraint imposed on the $\epsilon$ parameter by \cite{Alcaniz:2005dg}, only positive epsilon values are allowed. We fix the baryon content at the Planck Collaboration value $\Omega_{b}h^2=0.02226$. The radiation density parameter used is $\Omega_{r}=4.15 \times 10 ^{-5}h^{-2}$ and the photon density parameter is $\Omega_{\gamma}=2.469 \times 10 ^{-5}h^{-2}$ for a CMB temperature $T_{CMB}= 2.725$ K. 


The results of the analysis are presented in the Table \ref{tableresults}, and in Figures \ref{CPL5P} and \ref{BA5P} for the parameterisation P1 and P2, respectively. The figures show $1\sigma$ and $2\sigma$ confidence contours of the cosmological parameters and their posterior distribution marginalised over all other parameters. We also perform a statistical analysis using flat priors for the
entire set of parameters, whose results are presented in Table \ref{tableresults2}. We note that the bounds on $\epsilon$ are significantly reduced when the Gaussian prior is used, in agreement with the anticorrelation between $H_0$ and $\epsilon$ exhibited in Figs. \ref{CPL5P} and \ref{BA5P}.

In order to combine the observational and the thermodynamic constraints, we also perform a statistical analysis with fixed $\epsilon$ values. We analyse the cases with $\epsilon=0.002$ and $\epsilon=0.018$ for the parameterisation P1 and with $\epsilon=0.002$ and $\epsilon=0.020$ for the parameterisation P2, which correspond to the $1\sigma$ limits on  $\epsilon$ provided by our statistical analysis (see Table \ref{tableresults}).  In Figures \ref{CPLfig} and \ref{BAfig}, we present the new $1\sigma$ and $2\sigma$ confidence contours in the $w_0 - w_a$ plane and the thermodynamic constraints (\ref{c3}) - (\ref{cc4}) for the mentioned $\epsilon$ values. 
The redshift interval used in the thermodynamic constraints is $z \in (0.01,1.3) $,  which corresponds to the range of the nearest and farthest SNe of the sample, respectively. We find that the dark energy density satisfies the weak energy condition for this $z$ interval inside the plane region in Figures \ref{CPLfig} and \ref{BAfig}, therefore the second constraint set similar bounds to the ones derived in \cite{Silva12}.

We also find that the first thermodynamic constraint is sensitive to the values of the $\epsilon$ parameter. Figures \ref{CPLfig} and \ref{BAfig} show that in order to satisfy both thermodynamic conditions inside the $2\sigma$ confidence level, the value of $\epsilon$ must be very small. Indeed, it should be  smaller than the 1$\sigma$ upper limit allowed by the complete statistical analysis (see Table \ref{tableresults}). In the analysed cases with fixed $\epsilon$ values, we find an intersection region between both thermodynamic constraints and the $2\sigma$ observational confidence contour  delimited approximately by the triangles with vertices (for P1 and P2, respectively):

\begin{eqnarray}
\label{triangle}
(w_o,w_a) = \left\{
 \begin{tabular}{l}
\hspace{2.75cm} $\epsilon=0.002$\\
$(-0.99,-0.02),(-0.59,-0.71),(-0.68,-0.02)$\\  
\hspace{2.75cm} $\epsilon=0.018$\\
$(-0.89, -0.19),(-0.61, -0.68),(-0.67, -0.19)$
\quad\hspace{0.19cm}\\
\end{tabular}
\right. \nonumber
\end{eqnarray}

\begin{eqnarray}
\label{triangle}
(w_o,w_a) = \left\{
\begin{tabular}{l}
\hspace{2.75cm} $\epsilon=0.002$\\
$(-0.98, -0.02),(-0.65, -0.32),(-0.71, -0.02)$\\
\hspace{2.75cm} $\epsilon=0.020$\\
$(-0.83, -0.16),(-0.68, -0.29),(-0.70, -0.16)$\\
\quad\hspace{0.19cm}
\end{tabular}
\right. \nonumber
\end{eqnarray}

\section{Conclusions}

Relaxing the usual assumption of a minimal coupling between the components of the dark sector introduces significant changes in the predicted evolution of the universe. In this paper we have firstly discussed thermodynamic constraints on a class of interacting models assuming two parameterisations of the dark energy EoS (Eq. \ref{ps}). The constraints on $w$ come from the second law of thermodynamics and positiveness of entropy and are combined with current observational data through a Bayesian analysis. We have shown that this combination of physical and observational constraints on $w$ impose very tight limits on the $w_0 - w_a$ parametric space, as shown in Figs. \ref{CPLfig} and \ref{BAfig}. The thermodynamic analysis performed in this work generalises several cases previously discussed in the literature.

\section*{Acknowledgements}
JEG is supported by the DTI-PCI program of the Brazilian Ministry of Science, Technology, and Innovation (MCTI). RS is supported by the Conselho Nacional de Desenvolvimento Cient\'{\i}fico e tecnol\'ogico (CNPq). JSA acknowledges support from CNPq (grants  no. 310790/2014-0  and  400471/2014-0)  and FAPERJ (grant no.  204282).

\end{document}